\documentclass[twocolumn]{aastex62}

\usepackage{times,amsmath,amssymb,txfonts,graphicx,nicefrac,bm,xspace,color}
\usepackage{color,ulem,soul}

\renewcommand\emph[1]{\textit{#1}}

\definecolor{dark-red}{rgb}{0.75, 0.00, 0.00}
\definecolor{hlcolor}{rgb}{1.00, 0.90, 0.85}\sethlcolor{hlcolor}

\newcommand{\Sec}[1]{Sect.~\ref{sec:#1}}

\newcommand{\Fig}[1]{Fig.~\ref{fig:#1}}
\newcommand{\Figs}[2]{Figs.~\ref{fig:#1} and \ref{fig:#2}}
\newcommand{\Figure}[1]{Figure~\ref{fig:#1}}

\newcommand{\Table}[1]{Table~\ref{tab:#1}}

\newcommand{\Eqn}[1]{Eqn.~(\ref{eq:#1})}
\newcommand{\Equation}[1]{equation~(\ref{eq:#1})}
\newcommand{\Equations}[2]{equations~(\ref{eq:#1}) and (\ref{eq:#2})}

\newcommand\tms{\!\times\!}

\newcommand\xx{\hat{{\mathbf x}}}
\newcommand\yy{\hat{{\mathbf y}}}
\newcommand\zz{\hat{{\mathbf z}}}

\newcommand\U{\mathbf u}
\newcommand\V{\mathbf v}
\newcommand\B{\mathbf B}

\newcommand\mU{\mn{\mathbf{u}}}
\newcommand\mB{\mn{\mathbf{B}}}

\newcommand\EMF{\mbox{\boldmath{${\cal E}$}}}
\newcommand{\OO}{\bm{\Omega}}
\newcommand\Tf{\mathcal{B}}
\newcommand\mTf{\overline{\mathcal{B}}}
\newcommand\tfi{{(\mu)}}

\newcommand\tauc{\tau_{\rm c}}
\newcommand\etat{\eta_{\rm T}}

\newcommand{\mn}[1]{\overline{#1}}

\newcommand{\ee}[1]{$\,\times 10^{#1}$}

\renewcommand{\ij}{i\hspace{-0.75pt}j}

\newcommand{\simgt}%
           {\,\hbox{\lower0.35ex\hbox{$\sim$}\llap{\raise0.35ex\hbox{$>$}}}\,}
\newcommand{\simlt}%
           {\,\hbox{\lower0.35ex\hbox{$\sim$}\llap{\raise0.35ex\hbox{$<$}}}\,}

\newcommand\NIRVANA{\textsc{nirvana}\xspace}

\newcommand\aip{Leibniz-Institut f{\"u}r Astrophysik Potsdam (AIP),
  An der Sternwarte 16, 14482, Potsdam, Germany}

\newcommand\nbia{Niels Bohr International Academy, The Niels Bohr Institute,
  Blegdamsvej 17, DK-2100, Copenhagen \O, Denmark}

\received{December 16, 2021}\revised{January 1, 2022}\accepted{January 2, 2022}

\shorttitle{Non-instantaneous dynamo effects in MRI turbulence}
\shortauthors{Gressel \& Pessah}

\begin{document}

\title{\bf\large Finite-time response of dynamo mean-field effects
  in magnetorotational turbulence}

\correspondingauthor{Oliver Gressel}
\email{ogressel@aip.de}

\author[0000-0002-5398-9225]{Oliver Gressel} 
\affiliation{\aip}\affiliation{\nbia}

\author[0000-0001-8716-3563]{Martin E. Pessah} 
\affiliation{\nbia}


\begin{abstract}
  Accretion disc turbulence along with its effect on large-scale magnetic fields plays an important role in understanding disc evolution in general, and the launching of astrophysical jets in particular.
  Motivated by enabling a comprehensive sub-grid description for global long-term simulations of accretions discs, we aim to further characterize the transport coefficients emerging in local simulations of magnetorotational disc turbulence.
  For the current investigation, we leverage a time-dependent version of the test-field method, which is sensitive to the turbulent electromotive force (EMF) generated as a response to a set of pulsating background fields.
  We obtain Fourier spectra of the transport coefficients as a function of oscillation frequency. These are well approximated by a simple response function, describing a finite-time build-up of the EMF as a result of a time-variable mean magnetic field. For intermediate timescales (i.e., slightly above the orbital frequency), we observe a significant phase lag of the EMF compared to the causing field.
  Augmented with our previous result on a non-local closure relation in space, and incorporated into a suitable mean-field description that we briefly sketch out here, the new framework will allow to drop the restrictive assumption of scale separation.
\end{abstract}

\keywords{magnetic fields, turbulence, -- MHD -- methods: numerical -- }


\section{Introduction}
\label{sec:intro}

Thirty years after its \emph{ultimate} discovery by \citet{1991ApJ...376..214B}, the magnetorotational instability (MRI) is practically synonymous with accretion disc turbulence and is believed to be the key to understanding the structure and evolution of discs ranging from circumplanetary, to circumstellar (including those around black-holes or neutron stars, as well as the innermost and outer reaches of protoplanetary ones), all the way to active galactic nuclei.

Enhanced transport coefficients ---stemming from correlated fluctuations in the MRI turbulence--- have a profound impact on disc evolution. This can happen either directly, when disc accretion is enabled by angular momentum exchange through enhanced viscosity, or indirectly via a magnetocentrifugal disc outflow. Notably, the latter scenario requires large-scale ordered poloidal fields that are either created \emph{in situ} by a disc dynamo \citep[e.g.,][]{2003A&A...398..825V,2014ApJ...796...29S}, or ---when an inherited large-scale field is invoked--- are at least affected by enhanced field dissipation as a result of eddy diffusivity. Recent attempts of incorporating sub-grid-scale physics into jet-launching simulations \citep[e.g.][]{2013MNRAS.428...71B,2018ApJ...855..130F,2018MNRAS.477..127D,2020ApJ...900...59M,2020ApJ...900...60M,2021ApJ...911...85V} illustrate the need for comprehensive parametrizations that are ideally based on first-principles,  resolved MRI simulations.

Depending on (i) the level of inherited\,/\,accumulated net-vertical magnetic flux, and (ii) the relevance of the vertical disc structure, the MRI relies to a varying degree on the presence of an intrinsic dynamo of some sort to become a self sustained mechanism for powering disc accretion \citep[see][for an excellent review on this subject]{2019JPlPh..85d2001R}. A pronounced shortcoming of non-stratified box simulations is that they are very sensitive to the vertical aspect ratio \citep[see][]{2016MNRAS.456.2273S,2017MNRAS.470.2653W}.

Notably, when including vertical stratification, both local-box \citep[going back to][]{1995ApJ...446..741B} and global \citep[as recent as][]{2020MNRAS.494.4854D} fully non-linear MRI simulations alike robustly develop near-periodic cycles in the (horizontally\,/\,azimuthally averaged) mean magnetic field with a characteristic propagation away from the disc midplane -- providing a natural explanation to sustaining the MRI via a large-scale dynamo action \citep[e.g.][]{2005AN....326..787B,2008AN....329..725B,2010AN....331..101B}.

The morphology of this so-called butterfly diagram ---the hallmark of the mean-field dynamo--- was previously found to depend somewhat on the amount of net-vertical magnetic flux \citep[see, e.g.,][]{2015ApJ...810...59G,2016MNRAS.457..857S}. We here nevertheless focus on the limit of negligible net-vertical magnetic flux, which in a way is the crucial test for providing a robust accretion engine from MRI turbulence. Another important issue raised pertains to the onset of convective turnover \citep{2012ApJ...761..116B,2013ApJ...770..100G,2014ApJ...787....1H}, which was found to drastically affect the regularity of the magnetic-field cycles \citep[see the discussion in][]{2017MNRAS.467.2625C}. As with the net-vertical magnetic flux, we take a rather conservative stance and focus our investigation on the isothermal case, avoiding the complications associated with arguably more realistic thermodynamic representations.

A central question that remains unanswered is whether the dynamo wave can be reconciled with a conventional $\alpha\Omega$ dynamo (i.e., driven via the interplay of helical turbulence with differential rotation), and/or whether its dynamics are enforced by the near-exact conservation of magnetic helicity at high magnetic Reynolds number \citep[e.g.][]{2009ApJ...696.1021V,2010MNRAS.405...41G,2011ApJ...740...18O}. While the cycle period as a function of shear rate can nicely be explained using the dispersion relation of a near-critical $\alpha\Omega$~dynamo \citep{2015ApJ...810...59G}, the propagation direction \emph{away} from the midplane is still not well understood, possibly requiring a magnetic buoyancy contribution near the midplane \citep[][]{1998tbha.conf...61b}.

Both the spatial non-locality \citep[see][]{2002GApFD..96..319B} of the dynamo closure relation, the non-instantaneous aspects \citep[i.e., so-called ``memory effects'', e.g.,][]{2009ApJ...706..712H}, as well as their combined effect \citep[see, e.g.,][]{2012AN....333...71R} have been demonstrated to influence the characteristics of the dynamo cycle. Another comprehensive example of how finite-time effects can influence dynamo-generated fields has been presented by \citet{2013MNRAS.428.3569C,2013MNRAS.433.3274C} in the context of galactic magnetic fields. To complement our previous investigation of the scale-dependence\,/\,non-locality of the characteristic mean-field $\alpha$~effect in MRI turbulence \citep[][sect.~3.4]{2015ApJ...810...59G}, we here investigate the potential role of finite-time effects in the mean-field closure relation.

Our paper is organized in the following manner: Section~\ref{sec:methods} briefly describes the numerical simulations and introduces the newly adopted non-instantaneous closure relation to the mean-field induction equation, as well as how it can be captured using the test-field method. We present the results obtained from a fiducial MRI shearing-box simulation in Section~\ref{sec:results}, and we discuss how these findings may be exploited in the future, in Section~\ref{sec:discussion}.\vfill\null


\section{Methods}
\label{sec:methods}

As in previous work, we solve the equations of isothermal, ideal magnetohydrodynamics (MHD) in a local shearing-box \citep[e.g.][]{2007CoPhC.176..652G} frame of reference. Lacking explicit dissipation, the purist may call this an ``implicit'' large-eddy simulation (iLES). For practical purposes, we will nevertheless refer to these as direct numerical simulation (DNS). For brevity, we here only briefly recapitulate the essential properties of our numerical approach, and refer the reader to sections~2.1 et seqq. of \citet{2015ApJ...810...59G} for a more in-depth discussion, motivating our particular choices.

\subsection{Brief specification of the direct simulations} 
\label{sec:eqn}

We here use local Cartesian coordinates, ($x$, $y$, $z$), but refer to some tensor coefficients in cylindrical components, ($r$, $\phi$, $z$), for easier comparison with global models. Differential rotation is expressed via the parameter $q\equiv{\rm d}\ln\Omega/{\rm d}\ln r = -3/2$ for a Keplerian rotation curve, and we use the ``orbital advection'' scheme of \citet{2010ApJS..189..142S} to treat the background shear flow, $\V_{\rm K}\equiv q\,\Omega x\,\yy$, with the benefit of a position-independent truncation error.
The equations expressed in the local Eulerian velocity, $\V$, are

\begin{eqnarray}
      \frac{\partial\rho}{\partial t}
      + \nabla\cdot\left(\rho \V\right) & = & 0\,,
      \nonumber\\
      \frac{\partial\left(\rho\V\right)}{\partial t}
      + \nabla\cdot\left(\rho\mathbf{vv}
      + \mathrm{p}^{\star}\!
      - \mathbf{BB}\right) & = &
      - 2 \rho\ \Omega\ \zz\times\V \ -\rho\nabla\Phi\,,
      \nonumber\\
      \frac{\partial\mathbf{B}}{\partial t}
      - \nabla\times\left( \V\times\mathbf{B} \right) & = & 0\,,
      \label{eq:mhd}
\end{eqnarray}
with the total pressure $\mathrm{p}^{\star}\equiv \mathrm{p}+\mathbf{B}^2/2$, and the combined (i.e., tidal plus gravitational) effective potential
\begin{equation}
  \Phi(x,z) = q\,\Omega^2x^2 + \frac{1}{2}\Omega^2z^2\,,
\end{equation}
defined in the locally co-rotating frame of reference at fixed angular frequency $\OO\equiv\Omega\zz$.
Horizontal boundary conditions are shear-periodic \citep[see][for details]{2007CoPhC.176..652G}, and we apply standard outflow conditions in the vertical direction.
We chose an intermediate box size of $L_x\times L_y\times L_z = H\times\pi H\times 6H$ with a linear resolution of $\simeq 32/H$ in all space dimensions -- amounting to $32\times 100\times 192$ cells in the radial ($x$), azimuthal ($y$), and vertical ($z$) coordinate directions, respectively.
The initial plasma parameters in the disk midplane are $\beta_{\rm p}=800$ and $\beta_{\rm p}=2.2$\ee{5} for the zero-net-flux contribution, and the additional net-vertical field, respectively.
As previously, we include ~(i) an artificial mass diffusion term \citep[see][]{2011MNRAS.415.3291G} to circumvent undue time-step constraints resulting from low-density regions in the upper disc corona, and ~(ii) replenish the mass lost via outflow through the vertical domain boundary to obtain an overall steady-state disc structure.

\subsection{The non-instantaneous closure relation} 

Adopting the well established framework of mean-field magnetohydrodynamics \citep{1980opp..bookR....K}, we seek a parametrization for the turbulent  electromotive force, $\EMF \equiv \overline{\V'\tms \B'}$ with fluctuating magnetic and velocity fields defined as $\B'\equiv\B-\mB(z)$, and $\V'\equiv\V-\mn{\V}(z)$, respectively.\footnote{As we will be using the fluctuating velocity $\U \equiv \V - \V_{\rm K}$ in some places, we note that, because $\V_{\rm K}(x)$ vanishes when averaging, $\V'\equiv \U'$, trivially.} Here, and in the following, the overbar implies geometric averaging over horizontal slabs. This is the natural choice for the adopted box geometry and trivially satisfies the Reynolds rules required for a consistent mean-field description.
By virtue of its definition, the EMF captures correlations in fluctuating  velocity and magnetic field, whose non-zero mean appears as a source term on the right-hand-side of the (one-dimensional) mean-field induction equation
\begin{equation}
    \frac{\partial \mB(z)}{\partial t} - \nabla\times\left( \mn{\V}(z)\times\mB(z) \right) = \nabla\tms\EMF(z)\,.
  \label{eq:MF_ind}
\end{equation}
By construction, this equation describes the long-term evolution of the (comparatively slowly changing) \emph{mean} magnetic field under the effect of the underlying turbulence.
To make progress over a direct simulation approach, the EMF is then typically expanded into a linear functional of the mean magnetic field and its gradients as
\begin{equation}
  \EMF_i(z,t) = \alpha_{\ij}(z,t)\ \mn{B}_j(z,t)
          \ -\ \eta_{\ij}(z,t)\ \varepsilon_{\!jzl}\,\partial_z \mn{B}_l(z,t)\,,
          \label{eq:closure}
\end{equation}
where the indices $i,j,l$ label the coordinates $x,y$ and contraction over repeated indices is understood. Note that it is unnecessary to include the radial and azimuthal gradients in our case, which is due to the homogeneity of the turbulence in any given horizontal plane.

Under steady-state conditions, the second-rank tensors, $\alpha_{\ij}(z)$ and $\eta_{\ij}(z)$ become time-independent and are expected to capture the statistical properties of the chaotic flow \citep[see][]{1980opp..bookR....K}. If the system at hand is sufficiently anisotropic (e.g., due to rotation) and inhomogeneous (e.g., due to vertical gravity/stratification), $\alpha_{\ij}(z)$ ---as well as the off-diagonal elements of $\eta_{\ij}(z)$--- are expected to be non-vanishing. Together, the tensors encapsulate the emergence of the mean EMF as a response to imposing an external mean magnetic field --- or, in general, to the presence of a self-consistently evolving mean field. The purpose of the present paper is to elucidate a possible finite-time character of this response.

Typically, the turbulent closure coefficients are thought to connect $\EMF(z,t)$ to the mean magnetic field, $\mn{B}(z,t)$, and its curl $\varepsilon_{\!jzl}\,\partial_z \mn{B}_l(z,t)$, in a \emph{local} and \emph{instantaneous} fashion.\footnote{This local relation formally demands a ``scale separation'' between $\mn{\B}$, on one hand, and $\B'$, on the other hand (so that the slowly varying mean field can be pulled out of the integral describing the time evolution of the EMF).}
However, in contrast to this instantaneous characterization of the closure relation, the power-law nature of the turbulent cascade suggests that the space-time domain of dependence of $\EMF(z,t)$ is indeed finite -- implying so-called ``memory effects'' \citep{2009ApJ...706..712H}, that is, a delayed (i.e., out-of-phase) response to an applied mean field.\footnote{Note that, while we often speak of ``imposing'' or ``applying'' mean fields, and the EMF as a ``response'' (using the language of signal processing), these words can easily be replaced by ``pre-existing'' or ``emerging'' to better capture the spontaneous character of the chaotic turbulent flow.}
Under the assumption of statistically stationary turbulence, a simple non-instantaneous closure relation \citep[also see][]{2020MNRAS.494.1180G} can be formulated as a convolution integral in time of the form
\begin{eqnarray}
  \EMF_i(z,t) & = \int &
  \left[\;
  \hat{\alpha}_{\ij}(z,t')\,\mn{B}_j(z,t-t') \right. \nonumber \\ && - \left.
  \hat{\eta}_{\ij}(z,t')\ \varepsilon_{\!jzl}\,\partial_z\mn{B}_l(z,t-t')
  \;\right]
  \ {\rm d}t'\,.
  \label{eq:closure_conv_time}
\end{eqnarray}
In the local box geometry, the integral kernels $\hat{\alpha}_{\ij}(z,t')$ and $\hat{\eta}_{\ij}(z,t')$ are functions of the vertical coordinate, $z$, only. Moreover, in its Fourier-space representation, the above relation can be expressed as a simple multiplication \citep[see][appendix A]{2009ApJ...706..712H} with the Fourier transform, $\tilde{\alpha}_{\ij}$, of the kernel. That is (dropping the explicit z-dependence), we write
\begin{equation}
  \tilde{\EMF}_i(\omega) = \tilde{\alpha}_{\ij}(\omega)\ \tilde{\mn{B}_{\!j}}(\omega) \ -\ \tilde{\eta}_{\ij}(\omega)\ {\rm i} k_z \,\varepsilon_{\!jzl} \,\tilde{\mn{B}_l}(\omega) \,,
  \label{eq:closure_fourier_time}
\end{equation}
attributing a spectral flavor to the mean-field coefficients. While all of these quantities are in general complex functions, we can obtain real values (of, e.g., $\EMF_i$) in physical space and time by adding up the contributions from positive and negative frequencies.

Complementing the result of \citet{2015ApJ...810...59G} on non-local, scale-dependent character of the mean-field effects in magnetorotational turbulence, we here aim to obtain frequency-dependent closure coefficients, corresponding to convolution kernels in the time domain.
The frequency dependence can very naturally be obtained via the test-field (TF) method \citep{2005AN....326..245S,2007GApFD.101...81S} employing oscillating test fields as briefly outlined in the next section.

\subsection{The spectral test-field method} 

The defining advantage of the TF method, compared with other methods of inference, is that it relies on analytically prescribed ``test fields'', that can be chosen to span a non-degenerate basis for determining all tensor coefficients in an unambiguous manner. This differs from direct inversion methods \citep[see, e.g., discussion in][]{2020MNRAS.491.3870B}, that are founded on the (potentially degenerate) mean fields, $\mn{\B}(z,t)$, developing in the DNS.
To invert \Equation{closure_fourier_time}, and solve for the tensorial closure coefficients,  $\tilde{\alpha}_{\ij}(\omega)$ and $\tilde{\eta}_{\ij}(\omega)$, we apply the flavor of the method where the TFs, $\mTf_\tfi(z)$, are quadruplets of trigonometric functions \citep[also see, e.g.,][]{2005AN....326..787B,2008MNRAS.385L..15S,2008A&A...482..739B}:
\begin{eqnarray}
  \mn{\mathcal B}_{(0)} = \cos(\omega t)\, \cos(k_z z)\,\xx\,, &\quad
  \mn{\mathcal B}_{(1)} = \cos(\omega t)\, \sin(k_z z)\,\xx\,, &\nonumber\\
  \mn{\mathcal B}_{(2)} = \cos(\omega t)\, \cos(k_z z)\,\yy\,, &\quad
  \mn{\mathcal B}_{(3)} = \cos(\omega t)\, \sin(k_z z)\,\yy\,. &
  \label{eq:quadruplet}
\end{eqnarray}
For the purpose of determining the time-response, we here focus on a \emph{fixed} vertical scale $k_z=k_z^{\rm TF} = 2\pi/L_z$, with $L_z=6$ the vertical size of the box. This makes us sensitive to the coefficients representative of the largest scales available, and we refer the interested reader to section~3.4 of \citet{2015ApJ...810...59G} for a complementary discussion about the scale dependence \citep[also see][for the general case of full spatio-temporal dependence]{2012AN....333...71R}.
Having specified $k_z^{\rm TF}$, we moreover use eleven spectral modes $\omega=\omega^{\rm TF} = \nicefrac{1}{32}, \nicefrac{1}{16}, \nicefrac{1}{8}, \dots, 16, 32 \times 2\pi/P_0$, centered around $P_0=\nicefrac{1}{4}$ orbit for the temporal domain. We have arrived at this sampling interval by a combination of educated guessing and trial and error, and have found \emph{a posteriori} that the relevant dynamic range appears to be covered.

In total, we are hence solving $4\tms 11=44$ additional induction equations, one for each of the TF fluctuations, $\Tf'(\mathbf{r},t)$, alongside the DNS. In terms of the fluctuating velocity, $\U$, these are
\begin{equation}
  \partial_t\Tf' = \nabla \times \left[\;
      \U'\tms\mTf + (\mn{\U}+\V_{\rm K})\tms\Tf' - \!\overline{\U'\tms\Tf'} + \U'\tms\Tf' \;\right] \,.
  \label{eq:testfields}
\end{equation}
Importantly, these equations are passive in that they do not influence the evolving magnetic fields in the original DNS. Since we are dealing with MRI turbulence -- driven via an underlying genuinely magnetic instability -- there likely are pre-existing magnetic fluctuations, $\B'_0$, that are statistically independent from the developing (horizontal) mean fields. Such fluctuations, if correlated with the turbulent velocity, may result in an additional EMF, namely $\EMF_0\equiv\overline{\U'\tms \B'_0}$ -- which does, however, not enter our parametrization. To obtain the coefficients, we evaluate the corresponding mean electromotive force $\EMF^\tfi\equiv\overline{\V' \times \Tf'_\tfi}$ for each of the quadruplets from \Equation{quadruplet}. A formal solution to \Equation{closure_fourier_time} is then obtained as
\begin{equation}
  \left(\begin{array}{c}
      \tilde{\alpha}_{\ij}(k_z,\omega)\\[2pt]
      k_z\ \tilde{\beta}_{\ij z}(k_z,\omega)
    \end{array}\right) = {\rm e}^{\rm{i}\omega t}
  \left(\begin{array}{cc}
      \,\cos(k_z z) & \sin(k_z z) \\[2pt] \!-\sin(k_z z) & \cos(k_z z)
  \end{array}\right)
  \left(\begin{array}{c}
      \EMF_i^{(2j\!-\!2)} \\[2pt]
      \EMF_i^{(2j\!-\!1)}
    \end{array}\right)\,,
    \label{eq:tf_solution}
\end{equation}
where the tensors $\ \tilde{\eta}_{\ij}(k_z,\omega)\ $ (i.e., wrt. the current) and $\ \tilde{\beta}_{\ij z}(k_z,\omega)\ $ (i.e., wrt. the field gradients) are simply related via
\citep[see][]{2009ApJ...706..712H}
\begin{equation}
  \tilde{\eta}_{\ij} = \varepsilon_{jkz}\, \tilde{\beta}_{ikz}\,.
\end{equation}
In order to arrive to a statistical sound basis and eliminate by-chance fluctuations, \Equation{tf_solution} can simply be accumulated in time as needed.
This appears to be particularly relevant for slowly oscillating TFs, motivating our preference of long simulations time over grid resolution.
Note that in the time-dependent case, that is, for $\omega^{\rm TF}\!\ne\!0$, complex coefficients can arise, reflecting a frequency-dependent phase shift of the resulting EMF with respect to the originating oscillating TF. In practical terms, we replace the complex factor ${\rm e}^{\rm{i}\omega t}$ in \Equation{tf_solution} by either $\cos(\omega t)$ or $\sin(\omega t)$, in order to project the real\,/\,imaginary parts of the coefficients, respectively.
In the appendix of \citet{2020MNRAS.494.1180G}, we have benchmarked the implementation of the described spectral TF method, using the simple test case of helical forcing in the strictly kinematic limit.


\section{Results}
\label{sec:results}

We present results from a single generic shearing-box simulation of the MRI in the presence of a weak (i.e., midplane $\beta_{\rm p}=2.2$\ee{5}) net-vertical field, with $L_z=\pm 3\,H$ and outflow boundary conditions and a moderate resolution of 32 grid cells per pressure scale-height. The simulation quickly reaches a steady state with a dimensionless accretion stress (i.e., Reynolds + Maxwell) of about $0.01$, and we evolve the simulation for $555$ orbits to obtain decent statistics.

\begin{figure}
  \center\includegraphics[width=\columnwidth]{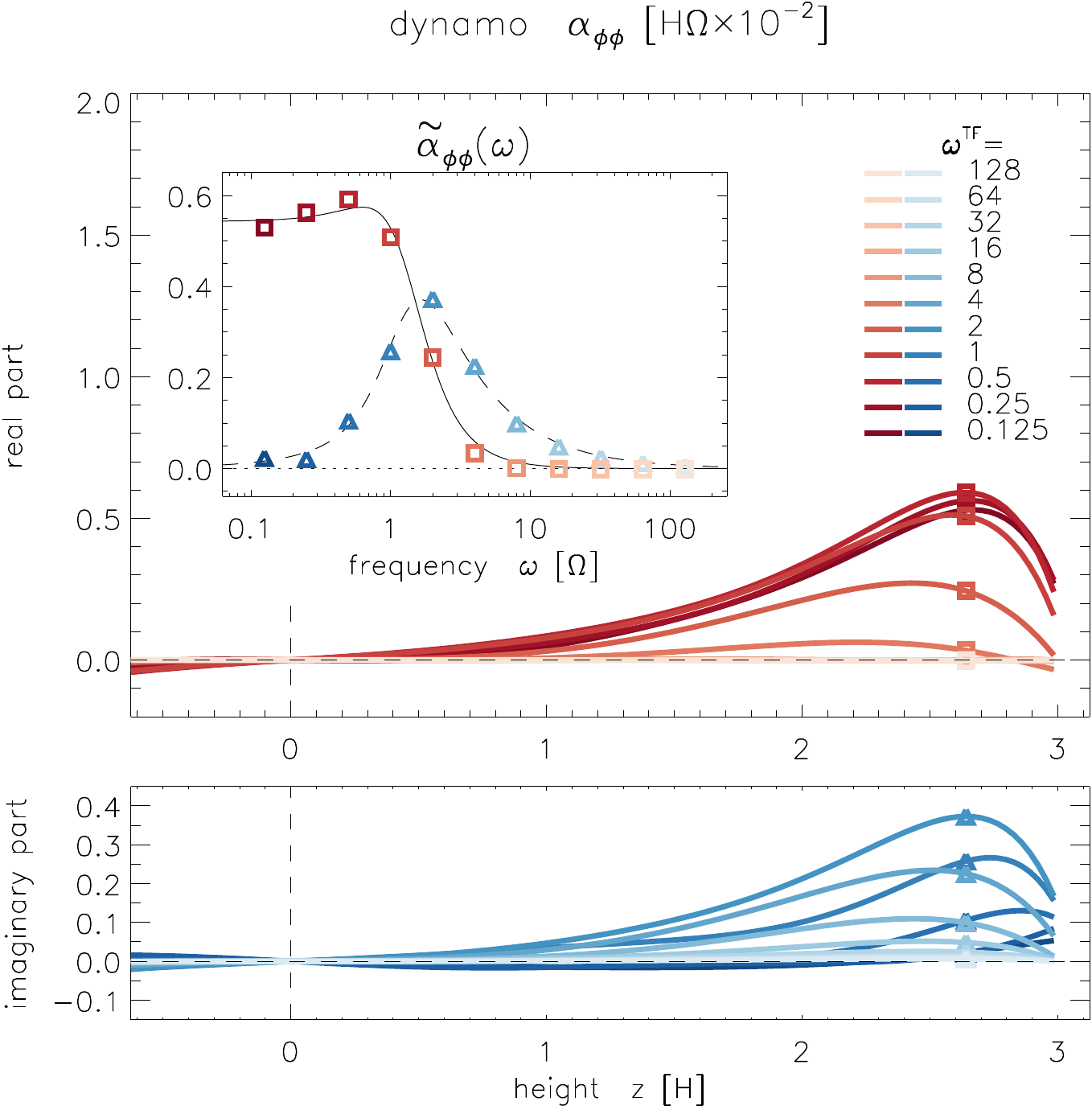}
  \caption{Profiles of the real part (upper panel$\,$/`$\square$') and imaginary part (lower panel$\,$/ `$\triangle$') of the  dynamo $\alpha_{\phi\phi}(z)$ coefficient as a function of height in the disc. Each curve represents one specific value of $\omega^{\rm TF}$, and the solution sampled at $z=2.67$ is plotted in the inset as a function of angular frequency, $\omega^{\rm TF}$, of the pulsating TFs, spanning three decades in dynamic range.}
  \label{fig:w_dynamo}
\end{figure}

In \Figure{w_dynamo}, we plot vertical profiles of the dynamo $\alpha$~effect, where we show real (upper panel/red tones) and imaginary part (lower panel/blue tones) separately.
The parametric curves represent the variation with the imposed oscillation frequency, $\omega^{\rm TF}$, of the TF inhomogeneity.\footnote{Note that these curves have been spatially filtered using a truncated series expansion into Legendre polynomials (up to order $l=12$), and this serves the purpose to extract a meaningful magnitude.}

Looking at the raw array of curves by eye makes it rather cumbersome to grasp anything but the most fundamental trends in the data.
As a remedy, and to illustrate the basic features of the spectral response, we resort to sampling point values at the arbitrary location $z=2.67$, roughly corresponding to the peak of the $\alpha$~profile.
The real (red$\,$/`$\square$') and imaginary (blue$\,$/`$\triangle$') amplitudes thus obtained are shown in the inset of \Fig{w_dynamo} along with best-fit response functions (solid/dashed, see \Sec{fits}, below, for details).

\begin{figure}
  \center\includegraphics[width=\columnwidth]{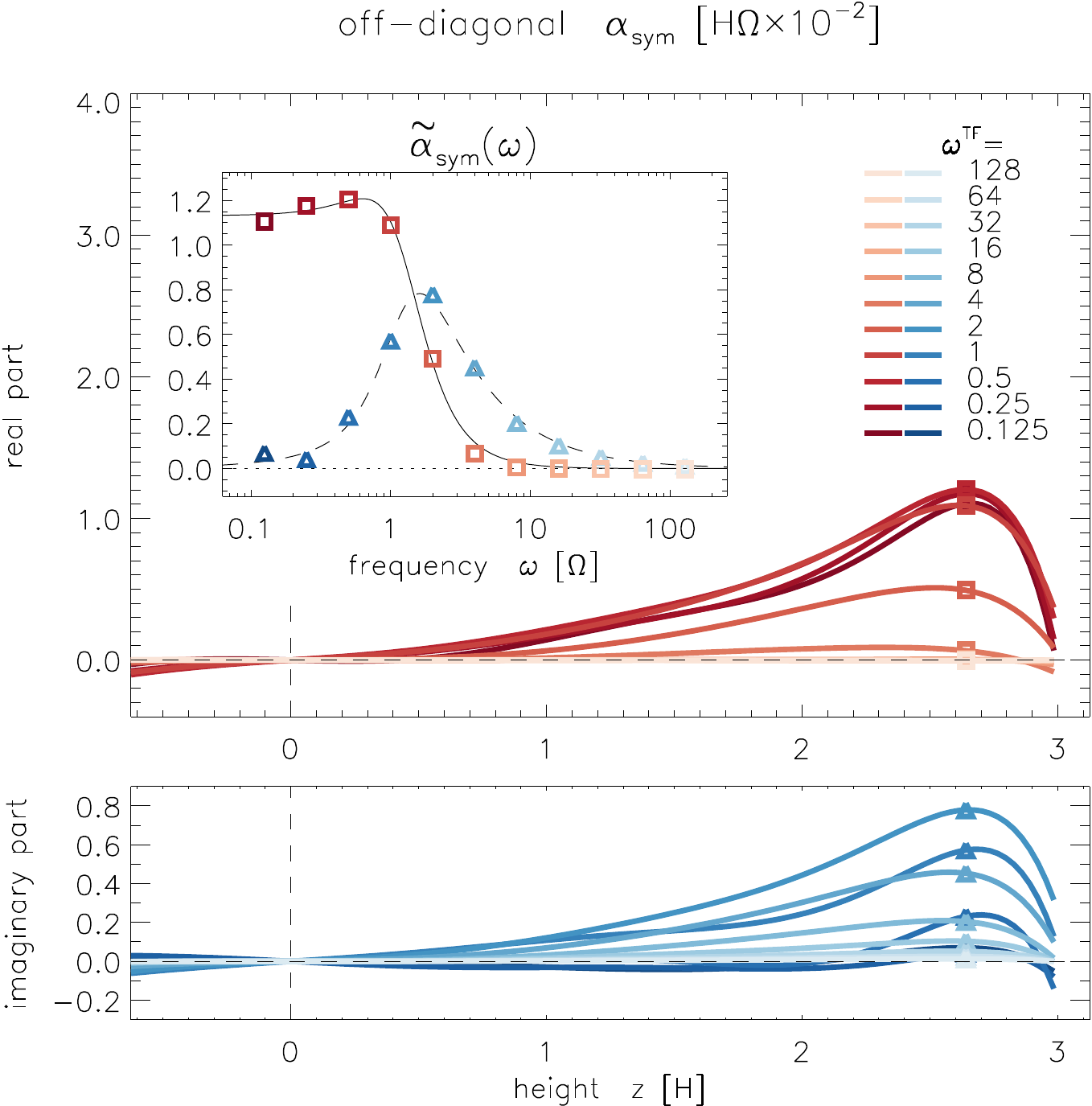}
  \caption{Same as \Figure{w_dynamo}, but for the \emph{symmetrized} off-diagonal element $\alpha_{\rm sym}\equiv \frac{1}{2}\,(\alpha_{\phi r} + \alpha_{r\phi})$ related to (differential) vertical turbulent pumping -- which we note is distinct from the conventional diamagnetic pumping related to the anti-symmetrized $\gamma \equiv \frac{1}{2}\,(\alpha_{\phi r} - \alpha_{r\phi})$.}
  \label{fig:w_offdia}
\end{figure}

We produce a corresponding plot ---shown in \Fig{w_offdia}--- for the symmetric off-diagonal components,  $\alpha_{\rm sym}\equiv \frac{1}{2}\,(\alpha_{\phi r} + \alpha_{r\phi})$, of the dynamo tensor.
One can see that apart from the overall amplitude, which is about a factor of two higher, these broadly match the characteristics of $\alpha_{\phi\phi}(z)$.

Unlike for the case of supernova-driven turbulence in the multi-phase interstellar medium \citep{2020MNRAS.494.1180G} ---where the off-diagonal elements were found to be anti-symmetric and distinct from the diagonal elements--- the diagonal  and off-diagonal components of the $\alpha$~tensor here (see \Figs{w_dynamo}{w_offdia}, respectively) show a rather similar time response.
The imaginary part displays a broad peak around $\omega=2$.
At the same time, the real part has a moderate overshoot around $\omega=1$, before it reaches the asymptotic value for slowly-varying mean fields.

The general shape of the response can be understood by visualizing the overall character of (rotating) turbulence.
Let us briefly recall what the $\alpha$~effect entails. It describes the emergence of a mean turbulent electromotive force as the direct consequence\footnote{That is, in a ``linear'' (or, leading-order) sense.} of the presence of a large-scale magnetic field.
In the limit of high frequencies, this ``presence'' obviously looses its coherent character and magnetic fluctuations created by the $\U'\tms\mTf$ term in \Eqn{testfields} will tend to become uncorrelated with the velocity and their contribution to the EMF will consequently average out to zero -- this is reflected in the vanishing amplitudes towards high frequencies.

Conversely, at low frequencies, we simply approach the previously reported \citep{2015ApJ...810...59G} amplitudes.
In particular, the imaginary part of the effect also tends to zero in this limit, so that the $\alpha$ effect becomes a real number, implying that there is no longer a phase difference.
The interesting regime falls in the region of intermediate frequencies, that roughly correspond to the eddy turnover time and/or rotational frequency of the turbulence.
Here the finite-time character of the relation between the imposed large-scale field $\mTf(t)$ (as a ``cause'') and the turbulent $\EMF(t)$ (as a ``response'') becomes most obvious.
In particular, owing to the non-instantaneous build-up of (correlated) magnetic fluctuations via the $\U'\tms\mTf$ term, the imaginary part deviates from zero, implying a phase-lag between cause and effect.
Since the $\alpha$~effect is thought to be related to the twisting-up of rising field loops by the rotation \citep[see, e.g.,][for an analytic calculation of the expected rotational dependence]{1993A&A...269..581R}, it appears natural that the effect peaks around $\omega\simeq\Omega$, where the pulsation of the assumed mean field matches the turnover of eddies that are affected by the Coriolis force.

\begin{figure}
  \center\includegraphics[width=\columnwidth]{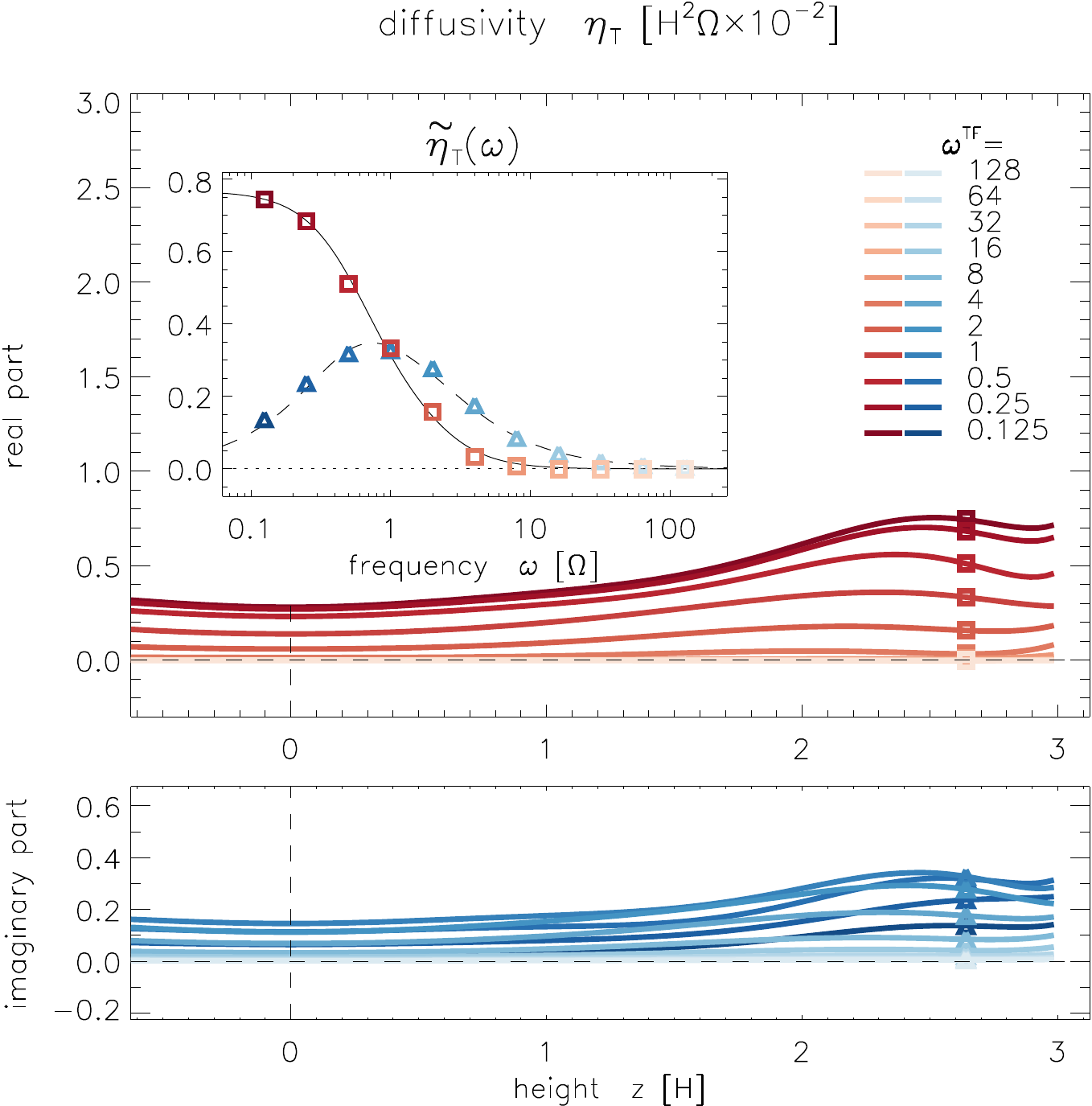}
  \caption{Same as \Figure{w_dynamo}, but for the turbulent magnetic diffusion coefficient $\etat \equiv \frac{1}{2}\,(\eta_{rr} + \eta_{\phi\phi})$ describing vertical diffusion of the field.}
  \label{fig:w_diff}
\end{figure}

In comparison with the $\alpha$~tensor, the diffusion coefficient, $\etat$, (shown in \Figure{w_diff}) shows a somewhat reduced coherence time. This broadly matches the expectation that the \emph{mixing aspect} of the chaotic flow field depends somewhat less on the buildup of correlated motions and is hence preserved further into the limit of high frequencies.
Moreover, the real part of $\etat(\omega)$ remains strictly monotonic around $\omega=1$, hinting at a reduced influence of the orbital timescale on the mere random diffusion of field.

\subsection{Characteristic response function} 
\label{sec:fits}

Pertaining to the non-instantaneous closure relation, and for the case of simple helically-forced turbulence, \citet{2009ApJ...706..712H} have demonstrated a frequency-dependence of the form of an ``oscillating decay'',
\begin{equation}
  \alpha(t) \ \propto\ \Theta(t)\ {\rm e}^{-t/\tauc}\,\cos(\omega_0 t)\,,
\end{equation}
where $\Theta(t)$ simply denotes the Heaviside step function, enforcing causality by suppressing dependence on future times.
Translated into Fourier space, the spectral shape function becomes
\begin{equation}
  \tilde{\alpha}(\omega) = A_0\;
  \frac{ 1-{\rm i}\,\omega\,\tauc}%
       { \left(1-{\rm i}\, \omega\,\tauc\right)^2 +
         \left(\omega_0\,\tauc\right)^2 }\,,
  \label{eq:osc_decay}
\end{equation}
with independent coefficients $A_0$, $\tauc$, and $\omega_0$ -- and with corresponding expressions for the other two coefficients of interest.
For the purpose of curve-fitting the frequency response, we write \Equation{osc_decay} separated into real and imaginary part as
\begin{eqnarray}
  \Re & = & A_0\;\frac{1 \,+\, (\omega^2+\omega_0^2)\,\tauc^2 }%
      { 4\,\omega^2\tauc^2
        + \left( 1 \,-\, (\omega^2-\omega_0^2)\,\tauc^2 \right)^2}\,,
  \label{eq:re_fit}
  \\[4pt]
  \Im & = & A_0\;\frac{1 \,+\, (\omega^2-\omega_0^2)\,\tauc^2 }%
      { 4\,\omega^2\tauc^2
        + \left( 1 \,-\, (\omega^2-\omega_0^2)\,\tauc^2 \right)^2}
      \;\omega\,\tauc\,,
  \label{eq:im_fit}
\end{eqnarray}
with separate sets of fit parameters $A_0$, $\tauc$, and $\omega_0$ for the three coefficients $\tilde{\alpha}_{\phi\phi}(\omega)$,  $\tilde{\alpha}_{\rm sym}(\omega)$, and $\tilde{\etat}(\omega)$, respectively.
We note that ---while we express the complex dependence in terms of two separate functional shapes via \Equations{re_fit}{im_fit}--- we fit the real and imaginary branches simultaneously in practical terms.

\begin{table}
  \begin{center}
  \begin{tabular}{llccccc}\hline\hline
    & & $A_0$ & $\tauc$ & $\omega_0$ & $\omega_0\,\tauc$ \\
    \hline
    dynamo        & $\alpha_{\phi\phi}$ & 0.88 & 3.92 & 0.20 & 0.78 \\
    off-diagonal  & $\alpha_{\rm sym}$  & 1.86 & 4.06 & 0.20 & 0.80 \\
    diffusivity   & $\etat$             & 0.56 & 3.05 & 0.17 & 0.52 \\
    \hline
  \end{tabular}
  \caption{Fit coefficients for the complex response function defined via \Equations{re_fit}{im_fit}, along with the derived parameter $\omega_0\,\tauc \simeq1$.\label{tab:fits}}
  \end{center}
\end{table}

The real part (solid lines) and imaginary parts (dashed lines) of the fitted curves are overlaid in the insets of Figures \ref{fig:w_dynamo}--\ref{fig:w_diff}, and represent the data rather well.
We, moreover, report best-fit values for coefficients (sampled at $z=2.67$) in \Table{fits}, where we also provide the dimensionless number, $\omega_0\tauc$, which we naively expect to be on the order of unity.

\begin{figure}
  \center\includegraphics[width=0.9\columnwidth]{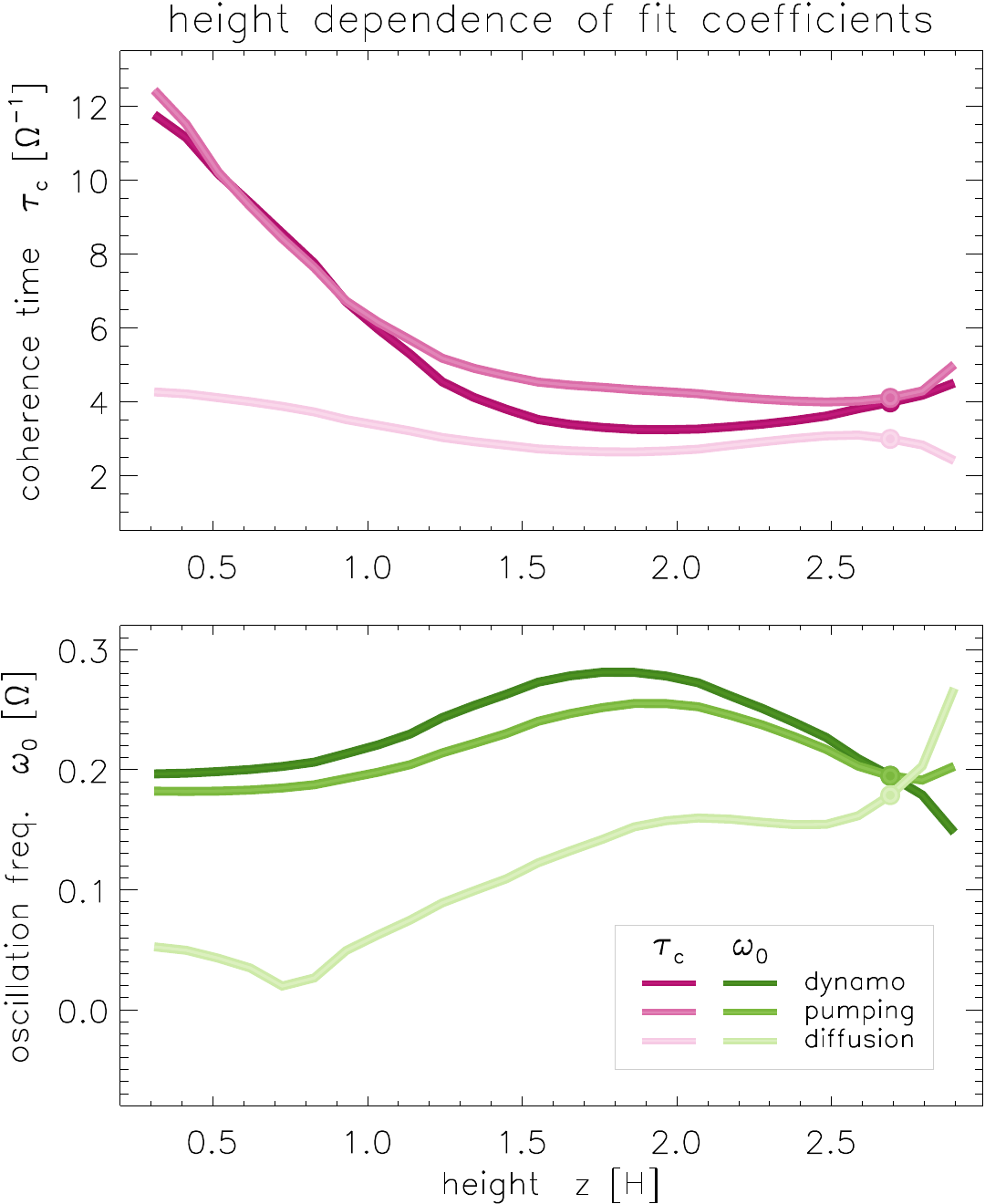}
  \caption{Height dependence of the two fit parameters from \Equation{osc_decay} describing the temporal behavior of the response. Circles indicate the values shown in \Table{fits} above. The fit amplitude, $A_0$, simply follows the original curves from the previous figures and is omitted here.}
  \label{fig:wdep_z}
\end{figure}

As a consequence of the vertical stratification in density, one may conjecture that the largest eddy--size depends on the height in the disc. This in turn should be reflected in the $\tauc$ and $\omega_0$ fit coefficients that we determine. To test this assumption, we plot the two numbers in the upper and lower panels of \Figure{wdep_z}, respectively. It can again be seen that the  dynamo $\alpha$ term and the off-diagonal elements are very similar, and at the same time, distinct from the diffusive coefficient. While $\tauc$ remains fairly constant for the latter, the former two show a pronounced increase (by a factor of three) of the turbulent correlation time towards the disc midplane. In contrast to this, their oscillatory parameter, $\omega_0$, remains fairly even in that limit but instead shows a moderate peak around $z=1.75\,H$. As consistent with the monotonic profile seen in \Fig{w_diff}, the $\omega_0$ parameter is reduced in the $\etat$ coefficient, and even drops to quite small values towards the disc midplane.

What precisely causes the observed trends is unclear at this point, and it is important to keep in mind that MRI turbulence is critically affected by magnetic forces so that intuition from hydrodynamic turbulence may be of limited value. Irrespective of this, comparing mean-field models with and without variation in $\tauc$ will allow to establish whether the seen variations do have an impact on the appearance of the butterfly diagram.


\section{Discussion \& Conclusions}
\label{sec:discussion}

The presented results clearly display the finite-time character of the mean-field dynamo effect emerging in stratified zero-net-flux MRI turbulence.
The approximate functional form presented in the preceding section will enable to incorporate the effects into a more comprehensive mean-field description of the evolution of large-scale magnetic fields in accretion discs.

As a first step in that direction, one may neglect the $\omega_0$ contribution, related to oscillatory behavior at intermediate frequencies.
In this case, \Equation{osc_decay} simplifies to
\begin{equation}
  \tilde{\alpha}(\omega) = A_0^{(\alpha)}\;\frac{1}{1-{\rm i}\, \omega\,\tauc^{(\alpha)}}\,,
\end{equation}
with corresponding expression for $\etat$. With the further approximation $\tauc^{(\alpha)}=\tauc^{(\eta)}$ (cf. \Table{fits} for judging to what degree this is justified), the characteristic time, $\tauc$, enters as a \emph{relaxation time} and implies a time-dependent (i.e., non-instantaneous) EMF response. In contrast to the algebraic relation from \Equation{closure}, this needs to be modeled as an extra PDE \citep[also see][]{2012AN....333...71R} of the form
\begin{equation}
  \big( 1 + \tauc\,\frac{\partial}{\partial t} \big)\; \EMF_i(z,t) \;=\;
  \alpha^{(0)}_{\ij}(z)\ \mn{B}_j(z,t)
  \ -\ \eta^{(0)}_{\ij}(z)\ \varepsilon_{\!jzl}\,\partial_z \mn{B}_l(z,t)\,,
\end{equation}
complementing the mean-field induction \Equation{MF_ind}, and where $\alpha^{(0)}_{\ij}(z)$ and $\eta^{(0)}_{\ij}(z)$ represent the above ``$A_0$'' coefficient (now measured for each tensor element individually). A very basic complete closure model would arise in combination with the non-locality derived in \citet{2015ApJ...810...59G}.
Expressed as a characteristic length, $l_{\rm c}$, relating to the finite ``domain of dependence'', the simplest form for the right-hand-side would then become
\begin{equation}
  \big( 1 + \tauc\,\frac{\partial}{\partial t}
  - l_{\rm c}^2 \frac{\partial}{\partial z^2} \big)\; \EMF_i(z,t) \;=\; \dots\,,
  \label{eq:MF_sophisticated}
\end{equation}
where the $l_c$ appears as a \emph{smoothing} term.

This, however, still neglects the possible advection (i.e., via a term $\mU\cdot\nabla\,\EMF_i$) of the EMF with the disc outflow, $\mn{u}_z$, as well as potential effects related to $\nabla\cdot\mU=\partial_z\,\mn{u}_z$. These may act to (de-)compress the EMF -- similar to the $(-\nabla\cdot\mU)\,\mB$ contribution to the $\nabla\times(\mU\times\mB)$ term in the induction equation itself. Moreover, if one were to restore the effect related to $\omega_0$, one would obtain a wave-like second-order time derivative of the EMF on the left-hand-side, as well as time derivative terms related to $\mB$ appearing on the right-hand side (M.~Rheinhardt, private communication).
A straightforward Cranck-Nicolson discretization of \Equation{MF_sophisticated} has already been implemented into a simple dynamo code.
In view of the mentioned complications, we however defer a detailed mean-field treatment along these lines to a later point in time.

Coming back to the inference of the dynamo coefficients from DNS, an often mentioned shortcoming of the TF method pertains to the absence of magnetic fluctuations stemming directly from the simulation. Operating in the so-called ``quasi-kinematic'' realm \citep[also see discussion in][]{2015ApJ...810...59G}, the QK-TFM is agnostic to the presence of a possible $\EMF_0$, and the velocity $\V(\mathbf{r},t)$ is the only manifest link in \Equation{testfields} to the physical evolution traced by the DNS.
While this may indeed be seen as a shortcoming of the current approach, we highlight that it merely implies that the detected mean-field effects likely are not exhaustive, but simply restricted to the chosen ansatz.
A promising avenue to accounting for these additional contributions has first been laid out by \citet{2010A&A...520A..28R} for a simplified set of equations (i.e., lacking the pressure and self-advection terms in the momentum equation). More recently, a workable solution has been found also for the complete set of MHD equations \citep*{2021arXiv210601107K}.
It appears natural to test this approach for MRI turbulence as well.


\section*{Acknowledgments}
We thank Tobias Heinemann for useful discussions, and Matthias Rheinhardt and Kandaswamy Subramanian for comments on a draft version. This work used the \NIRVANA code version 3.3, developed by Udo Ziegler at the Leibniz-Institut f{\"u}r Astrophysik Potsdam (AIP). All computations were performed on the \texttt{Steno} node at the Danish Center for Supercomputing (DCSC).



\clearpage
\begin{thebibliography}{}

\makeatletter
\relax

\def\mn@urlcharsother{\let\do\@makeother \do\$\do\&\do\#\do\^\do\_\do\%\do\~}
\def\mn@doi{\begingroup\mn@urlcharsother \@ifnextchar [ {\mn@doi@}
  {\mn@doi@[]}}
\def\mn@doi@[#1]#2{\def\@tempa{#1}\ifx\@tempa\@empty \href
  {http://dx.doi.org/#2} {doi:#2}\else \href {http://dx.doi.org/#2} {#1}\fi
  \endgroup}
\def\mn@eprint#1#2{\mn@eprint@#1:#2::\@nil}
\def\mn@eprint@arXiv#1{\href {http://arxiv.org/abs/#1} {{\tt arXiv:#1}}}
\def\mn@eprint@dblp#1{\href {http://dblp.uni-trier.de/rec/bibtex/#1.xml}
  {dblp:#1}}
\def\mn@eprint@#1:#2:#3:#4\@nil{\def\@tempa {#1}\def\@tempb {#2}\def\@tempc
  {#3}\ifx \@tempc \@empty \let \@tempc \@tempb \let \@tempb \@tempa \fi \ifx
  \@tempb \@empty \def\@tempb {arXiv}\fi \@ifundefined
  {mn@eprint@\@tempb}{\@tempb:\@tempc}{\expandafter \expandafter \csname
  mn@eprint@\@tempb\endcsname \expandafter{\@tempc}}}

\bibitem[\protect\citeauthoryear{{Balbus} \& {Hawley}}{{Balbus} \&
  {Hawley}}{1991}]{1991ApJ...376..214B}
{Balbus} S.~A.,  {Hawley} J.~F.,  1991, \mn@doi [\apj] {10.1086/170270}, \href
  {https://ui.adsabs.harvard.edu/abs/1991ApJ...376..214B} {376, 214}

\bibitem[\protect\citeauthoryear{{Bendre}, {Subramanian}, {Elstner}  \&
  {Gressel}}{{Bendre} et~al.}{2020}]{2020MNRAS.491.3870B}
{Bendre} A.~B.,  {Subramanian} K.,  {Elstner} D.,   {Gressel} O.,  2020,
  \mn@doi [\mnras] {10.1093/mnras/stz3267}, \href
  {https://ui.adsabs.harvard.edu/abs/2020MNRAS.491.3870B} {491, 3870}

\bibitem[\protect\citeauthoryear{{Blackman}}{{Blackman}}{2010}]{2010AN....331..101B}
{Blackman} E.~G.,  2010, \mn@doi [{\rm AN}] {10.1002/asna.200911304}, \href
  {https://ui.adsabs.harvard.edu/abs/2010AN....331..101B} {331, 101}

\bibitem[\protect\citeauthoryear{{Bodo}, {Cattaneo}, {Mignone}  \&
  {Rossi}}{{Bodo} et~al.}{2012}]{2012ApJ...761..116B}
{Bodo} G.,  {Cattaneo} F.,  {Mignone} A.,   {Rossi} P.,  2012, \mn@doi [\apj]
  {10.1088/0004-637X/761/2/116}, \href
  {https://ui.adsabs.harvard.edu/abs/2012ApJ...761..116B} {761, 116}

\bibitem[\protect\citeauthoryear{{Brandenburg}}{{Brandenburg}}{1998}]{1998tbha.conf...61b}
{Brandenburg} A.,  1998, in {Abramowicz} M.~A.,  {Bj{\"o}rnsson} G.,
  {Pringle} J.~E.,  eds, Theory of Black Hole Accretion Disks. pp 61--90

\bibitem[\protect\citeauthoryear{{Brandenburg}}{{Brandenburg}}{2005}]{2005AN....326..787B}
{Brandenburg} A.,  2005, \mn@doi [{\rm AN}] {10.1002/asna.200510414}, \href
  {https://ui.adsabs.harvard.edu/abs/2005AN....326..787B} {326, 787}

\bibitem[\protect\citeauthoryear{{Brandenburg}}{{Brandenburg}}{2008}]{2008AN....329..725B}
{Brandenburg} A.,  2008, \mn@doi [{\rm AN}] {10.1002/asna.200811027}, \href
  {https://ui.adsabs.harvard.edu/abs/2008AN....329..725B} {329, 725}

\bibitem[\protect\citeauthoryear{{Brandenburg} \& {Sokoloff}}{{Brandenburg} \&
  {Sokoloff}}{2002}]{2002GApFD..96..319B}
{Brandenburg} A.,  {Sokoloff} D.,  2002, \mn@doi [GAFD]
  {10.1080/03091920290032974}, \href
  {https://ui.adsabs.harvard.edu/abs/2002GApFD..96..319B} {96, 319}

\bibitem[\protect\citeauthoryear{{Brandenburg}, {Nordlund}, {Stein}  \&
  {Torkelsson}}{{Brandenburg} et~al.}{1995}]{1995ApJ...446..741B}
{Brandenburg} A.,  {Nordlund} A.,  {Stein} R.~F.,   {Torkelsson} U.,  1995,
  \mn@doi [\apj] {10.1086/175831}, \href
  {https://ui.adsabs.harvard.edu/abs/1995ApJ...446..741B} {446, 741}

\bibitem[\protect\citeauthoryear{{Brandenburg}, {R{\"a}dler}  \&
  {Schrinner}}{{Brandenburg} et~al.}{2008}]{2008A&A...482..739B}
{Brandenburg} A.,  {R{\"a}dler} K.~H.,   {Schrinner} M.,  2008, \mn@doi [\aap]
  {10.1051/0004-6361:200809365}, \href
  {https://ui.adsabs.harvard.edu/abs/2008A&A...482..739B} {482, 739}

\bibitem[\protect\citeauthoryear{{Bucciantini} \& {Del Zanna}}{{Bucciantini} \&
  {Del Zanna}}{2013}]{2013MNRAS.428...71B}
{Bucciantini} N.,  {Del Zanna} L.,  2013, \mn@doi [\mnras]
  {10.1093/mnras/sts005}, \href
  {https://ui.adsabs.harvard.edu/abs/2013MNRAS.428...71B} {428, 71}

\bibitem[\protect\citeauthoryear{{Chamandy}, {Subramanian}  \&
  {Shukurov}}{{Chamandy} et~al.}{2013a}]{2013MNRAS.428.3569C}
{Chamandy} L.,  {Subramanian} K.,   {Shukurov} A.,  2013a, \mn@doi [\mnras]
  {10.1093/mnras/sts297}, \href
  {https://ui.adsabs.harvard.edu/abs/2013MNRAS.428.3569C} {428, 3569}

\bibitem[\protect\citeauthoryear{{Chamandy}, {Subramanian}  \&
  {Shukurov}}{{Chamandy} et~al.}{2013b}]{2013MNRAS.433.3274C}
{Chamandy} L.,  {Subramanian} K.,   {Shukurov} A.,  2013b, \mn@doi [\mnras]
  {10.1093/mnras/stt967}, \href
  {https://ui.adsabs.harvard.edu/abs/2013MNRAS.433.3274C} {433, 3274}

\bibitem[\protect\citeauthoryear{{Coleman}, {Yerger}, {Blaes}, {Salvesen}  \&
  {Hirose}}{{Coleman} et~al.}{2017}]{2017MNRAS.467.2625C}
{Coleman} M. S.~B.,  {Yerger} E.,  {Blaes} O.,  {Salvesen} G.,   {Hirose} S.,
  2017, \mn@doi [\mnras] {10.1093/mnras/stx268}, \href
  {https://ui.adsabs.harvard.edu/abs/2017MNRAS.467.2625C} {467, 2625}

\bibitem[\protect\citeauthoryear{{Dhang}, {Bendre}, {Sharma}  \&
  {Subramanian}}{{Dhang} et~al.}{2020}]{2020MNRAS.494.4854D}
{Dhang} P.,  {Bendre} A.,  {Sharma} P.,   {Subramanian} K.,  2020, \mn@doi
  [\mnras] {10.1093/mnras/staa996}, \href
  {https://ui.adsabs.harvard.edu/abs/2020MNRAS.494.4854D} {494, 4854}

\bibitem[\protect\citeauthoryear{{Dyda}, {Lovelace}, {Ustyugova}, {Koldoba}  \&
  {Wasserman}}{{Dyda} et~al.}{2018}]{2018MNRAS.477..127D}
{Dyda} S.,  {Lovelace} R.~V.~E.,  {Ustyugova} G.~V.,  {Koldoba} A.~V.,
  {Wasserman} I.,  2018, \mn@doi [\mnras] {10.1093/mnras/sty614}, \href
  {https://ui.adsabs.harvard.edu/abs/2018MNRAS.477..127D} {477, 127}

\bibitem[\protect\citeauthoryear{{Fendt} \& {Ga{\ss}mann}}{{Fendt} \&
  {Ga{\ss}mann}}{2018}]{2018ApJ...855..130F}
{Fendt} C.,  {Ga{\ss}mann} D.,  2018, \mn@doi [\apj]
  {10.3847/1538-4357/aab14c}, \href
  {https://ui.adsabs.harvard.edu/abs/2018ApJ...855..130F} {855, 130}

\bibitem[\protect\citeauthoryear{{Gressel}}{{Gressel}}{2010}]{2010MNRAS.405...41G}
{Gressel} O.,  2010, \mn@doi [\mnras] {10.1111/j.1365-2966.2010.16440.x}, \href
  {https://ui.adsabs.harvard.edu/abs/2010MNRAS.405...41G} {405, 41}

\bibitem[\protect\citeauthoryear{{Gressel}}{{Gressel}}{2013}]{2013ApJ...770..100G}
{Gressel} O.,  2013, \mn@doi [\apj] {10.1088/0004-637X/770/2/100}, \href
  {https://ui.adsabs.harvard.edu/abs/2013ApJ...770..100G} {770, 100}

\bibitem[\protect\citeauthoryear{{Gressel} \& {Elstner}}{{Gressel} \&
  {Elstner}}{2020}]{2020MNRAS.494.1180G}
{Gressel} O.,  {Elstner} D.,  2020, \mn@doi [\mnras] {10.1093/mnras/staa663},
  \href {https://ui.adsabs.harvard.edu/abs/2020MNRAS.494.1180G} {494, 1180}

\bibitem[\protect\citeauthoryear{{Gressel} \& {Pessah}}{{Gressel} \&
  {Pessah}}{2015}]{2015ApJ...810...59G}
{Gressel} O.,  {Pessah} M.~E.,  2015, \mn@doi [\apj]
  {10.1088/0004-637X/810/1/59}, \href
  {https://ui.adsabs.harvard.edu/abs/2015ApJ...810...59G} {810, 59}

\bibitem[\protect\citeauthoryear{{Gressel} \& {Ziegler}}{{Gressel} \&
  {Ziegler}}{2007}]{2007CoPhC.176..652G}
{Gressel} O.,  {Ziegler} U.,  2007, \mn@doi [{\rm Comp. Phys. Comm.}]
  {10.1016/j.cpc.2007.01.010}, \href
  {https://ui.adsabs.harvard.edu/abs/2007CoPhC.176..652G} {176, 652}

\bibitem[\protect\citeauthoryear{{Gressel}, {Nelson}  \& {Turner}}{{Gressel}
  et~al.}{2011}]{2011MNRAS.415.3291G}
{Gressel} O.,  {Nelson} R.~P.,   {Turner} N.~J.,  2011, \mn@doi [\mnras]
  {10.1111/j.1365-2966.2011.18944.x}, \href
  {https://ui.adsabs.harvard.edu/abs/2011MNRAS.415.3291G} {415, 3291}

\bibitem[\protect\citeauthoryear{{Hirose}, {Blaes}, {Krolik}, {Coleman}  \&
  {Sano}}{{Hirose} et~al.}{2014}]{2014ApJ...787....1H}
{Hirose} S.,  {Blaes} O.,  {Krolik} J.~H.,  {Coleman} M. S.~B.,   {Sano} T.,
  2014, \mn@doi [\apj] {10.1088/0004-637X/787/1/1}, \href
  {https://ui.adsabs.harvard.edu/abs/2014ApJ...787....1H} {787, 1}

\bibitem[\protect\citeauthoryear{{Hubbard} \& {Brandenburg}}{{Hubbard} \&
  {Brandenburg}}{2009}]{2009ApJ...706..712H}
{Hubbard} A.,  {Brandenburg} A.,  2009, \mn@doi [\apj]
  {10.1088/0004-637X/706/1/712}, \href
  {https://ui.adsabs.harvard.edu/abs/2009ApJ...706..712H} {706, 712}

\bibitem[\protect\citeauthoryear{{K{\"a}pyl{\"a}}, {Rheinhardt}  \&
  {Brandenburg}}{{K{\"a}pyl{\"a}} et~al.}{2021}]{2021arXiv210601107K}
{K{\"a}pyl{\"a}} M.~J.,  {Rheinhardt} M.,   {Brandenburg} A.,  2021, arXiv
  e-prints, \href {https://ui.adsabs.harvard.edu/abs/2021arXiv210601107K} {p.
  arXiv:2106.01107}

\bibitem[\protect\citeauthoryear{{Krause} \& {Raedler}}{{Krause} \&
  {Raedler}}{1980}]{1980opp..bookR....K}
{Krause} F.,  {Raedler} K.~H.,  1980, {Mean-field magnetohydrodynamics and
  dynamo theory}

\bibitem[\protect\citeauthoryear{{Mattia} \& {Fendt}}{{Mattia} \&
  {Fendt}}{2020a}]{2020ApJ...900...59M}
{Mattia} G.,  {Fendt} C.,  2020a, \mn@doi [\apj] {10.3847/1538-4357/aba9d7},
  \href {https://ui.adsabs.harvard.edu/abs/2020ApJ...900...59M} {900, 59}

\bibitem[\protect\citeauthoryear{{Mattia} \& {Fendt}}{{Mattia} \&
  {Fendt}}{2020b}]{2020ApJ...900...60M}
{Mattia} G.,  {Fendt} C.,  2020b, \mn@doi [\apj] {10.3847/1538-4357/aba9d6},
  \href {https://ui.adsabs.harvard.edu/abs/2020ApJ...900...60M} {900, 60}

\bibitem[\protect\citeauthoryear{{Oishi} \& {Mac Low}}{{Oishi} \& {Mac
  Low}}{2011}]{2011ApJ...740...18O}
{Oishi} J.~S.,  {Mac Low} M.-M.,  2011, \mn@doi [\apj]
  {10.1088/0004-637X/740/1/18}, \href
  {https://ui.adsabs.harvard.edu/abs/2011ApJ...740...18O} {740, 18}

\bibitem[\protect\citeauthoryear{{Rheinhardt} \& {Brandenburg}}{{Rheinhardt} \&
  {Brandenburg}}{2010}]{2010A&A...520A..28R}
{Rheinhardt} M.,  {Brandenburg} A.,  2010, \mn@doi [\aap]
  {10.1051/0004-6361/201014700}, \href
  {https://ui.adsabs.harvard.edu/abs/2010A&A...520A..28R} {520, A28}

\bibitem[\protect\citeauthoryear{{Rheinhardt} \& {Brandenburg}}{{Rheinhardt} \&
  {Brandenburg}}{2012}]{2012AN....333...71R}
{Rheinhardt} M.,  {Brandenburg} A.,  2012, \mn@doi [{\rm AN}]
  {10.1002/asna.201111625}, \href
  {https://ui.adsabs.harvard.edu/abs/2012AN....333...71R} {333, 71}

\bibitem[\protect\citeauthoryear{{Rincon}}{{Rincon}}{2019}]{2019JPlPh..85d2001R}
{Rincon} F.,  2019, \mn@doi [Journal of Plasma Physics]
  {10.1017/S0022377819000539}, \href
  {https://ui.adsabs.harvard.edu/abs/2019JPlPh..85d2001R} {85, 205850401}

\bibitem[\protect\citeauthoryear{{R{\"u}diger} \& {Kitchatinov}}{{R{\"u}diger}
  \& {Kitchatinov}}{1993}]{1993A&A...269..581R}
{R{\"u}diger} G.,  {Kitchatinov} L.~L.,  1993, \aap, 269, 581

\bibitem[\protect\citeauthoryear{{Salvesen}, {Simon}, {Armitage}  \&
  {Begelman}}{{Salvesen} et~al.}{2016}]{2016MNRAS.457..857S}
{Salvesen} G.,  {Simon} J.~B.,  {Armitage} P.~J.,   {Begelman} M.~C.,  2016,
  \mn@doi [\mnras] {10.1093/mnras/stw029}, \href
  {https://ui.adsabs.harvard.edu/abs/2016MNRAS.457..857S} {457, 857}

\bibitem[\protect\citeauthoryear{{Schrinner}, {R{\"a}dler}, {Schmitt},
  {Rheinhardt}  \& {Christensen}}{{Schrinner}
  et~al.}{2005}]{2005AN....326..245S}
{Schrinner} M.,  {R{\"a}dler} K.~H.,  {Schmitt} D.,  {Rheinhardt} M.,
  {Christensen} U.,  2005, \mn@doi [{\rm AN}] {10.1002/asna.200410384}, \href
  {https://ui.adsabs.harvard.edu/abs/2005AN....326..245S} {326, 245}

\bibitem[\protect\citeauthoryear{{Schrinner}, {R{\"a}dler}, {Schmitt},
  {Rheinhardt}  \& {Christensen}}{{Schrinner}
  et~al.}{2007}]{2007GApFD.101...81S}
{Schrinner} M.,  {R{\"a}dler} K.-H.,  {Schmitt} D.,  {Rheinhardt} M.,
  {Christensen} U.~R.,  2007, \mn@doi [GAFD] {10.1080/03091920701345707}, \href
  {https://ui.adsabs.harvard.edu/abs/2007GApFD.101...81S} {101, 81}

\bibitem[\protect\citeauthoryear{{Shi}, {Stone}  \& {Huang}}{{Shi}
  et~al.}{2016}]{2016MNRAS.456.2273S}
{Shi} J.-M.,  {Stone} J.~M.,   {Huang} C.~X.,  2016, \mn@doi [\mnras]
  {10.1093/mnras/stv2815}, \href
  {https://ui.adsabs.harvard.edu/abs/2016MNRAS.456.2273S} {456, 2273}

\bibitem[\protect\citeauthoryear{{Stepanovs}, {Fendt}  \&
  {Sheikhnezami}}{{Stepanovs} et~al.}{2014}]{2014ApJ...796...29S}
{Stepanovs} D.,  {Fendt} C.,   {Sheikhnezami} S.,  2014, \mn@doi [\apj]
  {10.1088/0004-637X/796/1/29}, \href
  {https://ui.adsabs.harvard.edu/abs/2014ApJ...796...29S} {796, 29}

\bibitem[\protect\citeauthoryear{{Stone} \& {Gardiner}}{{Stone} \&
  {Gardiner}}{2010}]{2010ApJS..189..142S}
{Stone} J.~M.,  {Gardiner} T.~A.,  2010, \mn@doi [\apjs]
  {10.1088/0067-0049/189/1/142}, \href
  {https://ui.adsabs.harvard.edu/abs/2010ApJS..189..142S} {189, 142}

\bibitem[\protect\citeauthoryear{{Sur}, {Brandenburg}  \& {Subramanian}}{{Sur}
  et~al.}{2008}]{2008MNRAS.385L..15S}
{Sur} S.,  {Brandenburg} A.,   {Subramanian} K.,  2008, \mn@doi [\mnras]
  {10.1111/j.1745-3933.2008.00423.x}, \href
  {https://ui.adsabs.harvard.edu/abs/2008MNRAS.385L..15S} {385, L15}

\bibitem[\protect\citeauthoryear{{Vishniac}}{{Vishniac}}{2009}]{2009ApJ...696.1021V}
{Vishniac} E.~T.,  2009, \mn@doi [\apj] {10.1088/0004-637X/696/1/1021}, \href
  {https://ui.adsabs.harvard.edu/abs/2009ApJ...696.1021V} {696, 1021}

\bibitem[\protect\citeauthoryear{{Vourellis} \& {Fendt}}{{Vourellis} \&
  {Fendt}}{2021}]{2021ApJ...911...85V}
{Vourellis} C.,  {Fendt} C.,  2021, \mn@doi [\apj] {10.3847/1538-4357/abe93b},
  \href {https://ui.adsabs.harvard.edu/abs/2021ApJ...911...85V} {911, 85}

\bibitem[\protect\citeauthoryear{{Walker} \& {Boldyrev}}{{Walker} \&
  {Boldyrev}}{2017}]{2017MNRAS.470.2653W}
{Walker} J.,  {Boldyrev} S.,  2017, \mn@doi [\mnras] {10.1093/mnras/stx1032},
  \href {https://ui.adsabs.harvard.edu/abs/2017MNRAS.470.2653W} {470, 2653}

\bibitem[\protect\citeauthoryear{{von Rekowski}, {Brandenburg}, {Dobler},
  {Dobler}  \& {Shukurov}}{{von Rekowski} et~al.}{2003}]{2003A&A...398..825V}
{von Rekowski} B.,  {Brandenburg} A.,  {Dobler} W.,  {Dobler} W.,   {Shukurov}
  A.,  2003, \mn@doi [\aap] {10.1051/0004-6361:20021699}, \href
  {https://ui.adsabs.harvard.edu/abs/2003A&A...398..825V} {398, 825}

\makeatother

\end{thebibliography}


\end{document}